# ESTIMATION AND COMPENSATION OF INTER CARRIER INTERFERENCE IN WiMAX PHYSICAL LAYER UNDER VARIOUS CHANNEL MODELS

Tarun Kumar Juluru[1] and Anitha sheela Kankacharla[2]

[1] Sumathi Reddy Institute of Technology For Women, Warangal,
Andhra Pradesh, India.
tarunjuluru@yahoo.com
[2] Department of ECE, JNTU College of Engineering, Hyderabad,
Andhra Pradesh, India.
kanithasheela@gmail.com

## ABSTRACT

*WiMAX is Wireless Interoperability for Microwave Access has emerged as a promising solution for transmission of higher data rates for fixed and mobile applications. IEEE 802.16d and e are the standards proposed by WiMAX group for fixed and mobile. As the wireless channel have so many limitation Such as Multipath, Doppler spread, Delay spread and Line Of Sight (LOS)/Non Line Of Sight (NLOS) components. To attain higher data rates the Multi Carrier System with Multiple Input and Multiple Output (MIMO) is incorporated in the WiMAX. The Orthogonal Frequency Division Multiplexing (OFDM) is a multi carrier technique used with the WiMAX systems. In OFDM the available spectrum is split into numerous narrow band channels of dissimilar frequencies to achieve high data rate in a multi path fading environment. And all these sub carriers are considered to be orthogonal to each other. As the number of sub carriers is increased there is no guarantee of sustained orthogonality, i.e. at some point the carriers are not independent to each other, and hence where the orthogonality can be loosed which leads to interference and also owing to the synchronization between transmitter and receiver local oscillator, it causes interference known as Inter Carrier Interference (ICI). The systems uses MIMO-OFDM will suffer with the effects of ICI and Carrier Frequency Offset (CFO) "ε". However these affect the power leakage in the midst of sub carriers, consequently degrading the system performance. In this paper a new approach is proposed in order to reduce the ICI caused in WiMAX and improve the system performance. In this scheme at the transmitter side the modulated data and a few predefined pilot symbols are mapped onto the non neighboring sub carriers with weighting coefficients of +1 and -1. With the aid of pilot symbols the frequency offset is exactly estimated by using Maximum Likelihood Estimation (MLE) and hence can be minimized. At demodulation stage the received signals are linearly combined along with their weighted coefficients and pilot symbols, called as Pilot Aided Self Cancellation Method (PASCS). And also to realize the various wireless environments the simulations are carried out on Stanford University Interim (SUI) channels. The simulation results shows that by incorporating this method into WiMAX systems it performs better when the Line Of Sight (LOS) component is present in the transmission and also it improves the Bit Error Rate (BER) and Carrier to Interference Ratio (CIR). The CIR can be improved 20 dB. In this paper the effectiveness of PASCS scheme is compared with the Self Cancellation Method (SCM). It provides accurate estimation of frequency offset and when residual CFO is less significant the ICI can be diminished successfully.*

# 1. INTRODUCTION

Worldwide interoperability for microwave access IEEE802.16 is a standard proposed by WiMAX group. 802.16d is for fixed networks and e for mobile broadband networks. The physical layer of WiMAX is characterized by various advanced technologies [1] such as multi carrier schemes, higher order modulation, forward error correction, Adaptive Modulation and Coding (AMC) and MIMO to support the higher data rate transmission in multi path radio environment.

The physical layer of 802.16d is based on LOS and 802.16e is based on NLOS. Hence in the 802.16e the Physical layer and MAC layer is defined. The OFDM is the promising technique for achieving higher data rates with maximum utilization of band width in multi path environment, Indeed Multiple Input and Multiple Output (MIMO) combined with OFDM got better response in mobile WiMAX to realize higher bit rate and reliable transmission.

The Physical layer of WiMAX is also employed with other features such as Adaptive Modulation and Coding for abstraction or to realize the link, Hybrid Automatic Repeat and Request (HARQ) to enhance the capacity of mobile applications. In addition to this it also employs advanced channel coding schemes such as Convolutional Turbo Coding (CTC), Low Density Parity Check coding (LDPC) for forward error correction and acquire the Shannon's limit. In AMC it supports up to 64-QAM with a code rate of 5/6 [2].

The systems using OFDM ruthlessly suffers to sustain the orthogonality between subcarriers [3]. Also, any frequency offset between transmitter and receiver, if not corrected, will result in blurring of the information between the subcarriers. This is called inter-carrier interference (ICI), and is caused by a loss in the orthogonality of the carriers. In the worst case, in which the offset is of the order of the sub-carrier spacing, the information will mostly land in the adjacent subcarrier, causing all the bits on that carrier to be lost [5].In this paper a modern approach is proposed to compensate the ICI under WiMAX environment.

The paper is organized as follows. In Section II, Mobile WiMAX system model is discussed. Next, proposed model to compensate ICI and Simulate for various channel models are explained in Section III. The simulations for various parameters are discussed in Section IV. Finally, Section V concludes this paper

# 2. SYSTEM MODEL

IEEE 802.16 is a standard for broad band wireless access (BWA) air interface specification for the wireless metropolitan area network. WiMAX architecture is mainly considers fixed and mobile that uses point to multi point communication.

The physical layer of WiMAX will process the data frames received from upper layers and send to the receiver through a wireless channel in suitable format with 99.999percent reliability [4]. To achieve these, various stages of physical layer are configured as Forward Error Correction (FEC), Modulation, MIMO encoder and mapping to OFDM in the transmitter stage and similar reverse stages at the receiver.

## .2.1. Forward Error Correction and Coding

In the transmission of digital signals the FEC block of Fig.1. is a class of signal transformation designed to improve the performance. It enables the transmitted signal to better with stand the effects of various channel impairments such as noise, fading and interference. These coding schemes introduce redundancies in the transmitted signals [4], so that they may be exploiting at the receiver. These codes are operated at various code rates and which is defined as the ratio of generated bits at the output 'n' and data bits at the input "k" in our simulations the convolution codes are used.

## 2.2. Interleaving

The wireless   channel exhibits the multipath fading where signal arrives at the receiver over two or more paths. Due to this the received signals are out of phase with each other and hence results in distortion at the receiver. To avoid such problems interleaving is done.   Interleaver is a device that arranges the order of sequences of symbols in a deterministic manner. To minimize the bit correlation the size of interleaver plays a major role in the performance which may increase the delay. In this paper the interleaver is selected in such a way that the data is written in column format and read in row wise.

## 2.3. Modulation schemes

Modulation is the process by which the signal is transformed in to waveforms that are compatible with the characteristics of the channel the WiMAX scheme uses higher order modulation. In this paper the digital modulation schemes used are QPSK,8-QAM and 16-QAM.

## 2.4. Pilot Insertion

Pilot is basically a reference carrier or a signal which is known at the receiver used for channel estimation. There are many different methods for pilot insertion and transmission of pilots. In OFDM there are two kinds of pilot insertion. Block insertion and comb type. In comb method special pilot symbols are assigned to few pilot carriers dedicately, and pilot carriers are spaced equally along with the data sub carriers. In block type each OFDM symbol block assigned to a sub carrier is having some time duration for sending training sequence. The spacing of pilots in time domain and frequency domain are depends on delay spread $\tau$max at sub carrier spacing, Doppler spread and symbol time Ts.

$$Npt \cong \frac{1}{nn.\frac{1}{1}*}.$$
$$Npf \cong \frac{1}{\Delta f Tmax}$$

Where the $N_{pt}$ and $N_{pf}$ are the maximum number of subcarriers placed in Time and Frequency domain

## 2.5. MIMO encoder

The space time encoder as shown in Fig.1 provides a mapping between symbols in to an Nt dimensional stream. The output stream is transmitted through the antennas to reduce the noise and the mapping process depends on the MIMO method used in the system. The 802.16e standard MIMO is used in the simulation. The WiMAX system uses various MIMO such as 2x1 almouti's scheme,2x2 and 4x4 etc.

## 2.6. Mapping to OFDM

Orthogonal Frequency Division Multiplexing is used to maintain the orthogonality between the sub carriers, and this can be  achieve by applying IFFT to the data carriers generated from MIMO encoders. The OFDM sub carriers are basically in frequency  domain hence by using IFFT the orthogonality between the sub carriers and  can be maintained. In this paper the size of FFT is selected as per the standard of IEEE802.16e WiMAX group [2], as the more size of FFT increase the resolution of the system.

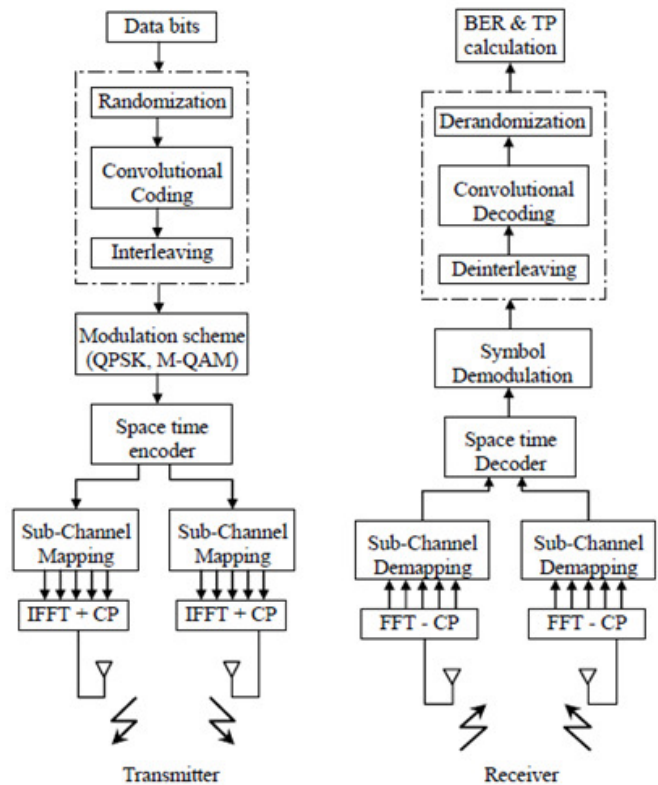

Figure 1.  Standard WiMAX system IEEE 802.16d

## 2.7. Cyclic prefix

Inter Symbol Interference is the major problem in high data rate communications. To avoid this a guard period is inserted between the transmitted symbols. The most effective way of using guard period is adding cyclic prefix to the symbol. Transmitting the cyclic prefix of data during guard interval results in converting linearly convolution of channel into circular convolute channel with transmitted signal. When a guard interval is longer than the channel impulse response or the multi path delay then ISI is eliminated.

## 2.8. Channel models

The channel is communication medium between transmitter and the receiver. The wireless signal changes its characteristics as it travels to certain distance from transmitter to receiver. These characteristics depends on path taken by the signal, distance between the transmitter and receiver and environment around the path. The symbol received at the receiver is results of convolution of the transmitted symbol is multiplied with channel response and added with white noise.

## 3. PROPOSED MODEL FOR ICI CANCELLATION IN WiMAX SYSTEMS

In the Fig.2 two more blocks are added to the WiMAX architecture, these are ICI cancelling modulation at the transmitter and demodulation at the receiver. In the transmitter side the data symbols and pilot symbols are mapped on to non neighboring sub carriers 'r' and 'N-r+1'.These subcarriers are multiplied with ICI cancelling coefficients +1 and -1 ,so that the sub carriers become $X_{i,N} = -X_{i,1}$, $X_{i,N-1} = -X_{i,2}$ ……….etc. and these modulated data symbols and pilot symbols are fed in to the input of space time encoder.

The space time encoding is a technique to support the MIMO system to maintain the diversity in STBC technique multiple copies of data are transmitted in different time slots across number of antennas. In this paper the almouti's space time coding with 2-transmit and 1 receiving antenna are used.

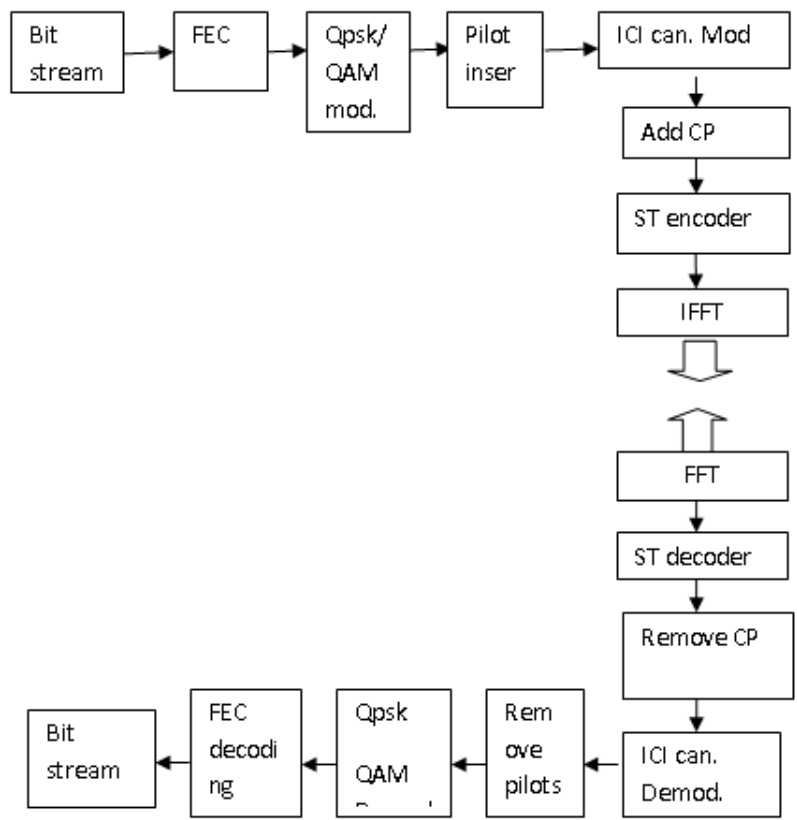

Figure 2.  Proposed WiMAX system with ICI cancellation

Let $X_{i,1}$ and $X_{i,2}$ are sent by the two transmitting antenna 1,2 in the time slot 1  $X^{*}_{i,1}$ and $X^{*}_{i,2}$ is sent by antenna 1,2 in the time slot 2. In space time encoding the complex conjugate signals are transmitted to full fill the orthogonality. If channel parameters are denoted by $h_1$,$h_2$ and the received vector $r_1$and $r_2$ by the receiving antenna in time slot 1 and 2 .

$$r_1 = h_1 s_1 + h_2 s_2 + n_1. \tag{1}$$

$$r_2 = s_2^* h_1 - s_1^* h_2 + n_2 \qquad (2)$$

.

Where n1 and n2 are Gaussian noise at time slot T1 and T2.

And these signals are combined before forwarding to the receiver. Then the combined signals are given as

$$\widetilde{s_1} = h_1^* n_1 + h_2 r_2^* . \qquad (3)$$

$$\widetilde{s_2} = h_2^* n_1 + h_1 r_2^* . \qquad (4)$$

Substituting the eqn(1)&(2) in eqn(3)&(4)

$$\widetilde{s_1} = (\alpha_1^2 + \alpha_2^2)s_1 + h_1^* n_1 + h_2 n_2^* . \qquad (5)$$

$$\widetilde{s_2} = (\alpha_1^2 + \alpha_2^2)s_2 - h_1^* n_2 + h_2 n_1^* \qquad (6)$$

The $\alpha_1^2$ is the squared magnitude of channel transfer function $h_i$. The Maximum likely hood (ML) decoder will estimate the transmitted signal from the signals $\widetilde{s_1}$ and $\widetilde{s_2}$ and decodes the signal $s_1$ and $s_2$ by linear processing of all possible values of $s_1$ and $s_2$ such that the eqn (5) and (6) are minimized,

$$s_1 = \arg\min |(h_1^* n_1 + h_2 r_2^*) - s_1|^2 + A|s_1|^2 . \qquad (7)$$

$$s_2 = arg\min |(h_2^* n_1 - h_1 r_2^*) - s_2|^2 + A|s_2|^2 . \qquad (8)$$

Where the A is given as

$$A = -1 + \sum_i h_i . \qquad (9)$$

Then the transmitted symbols after decoding will be given to the input of ICI cancelling modulation block in which the ICI is eliminated.

At the transmitter side the IFFT is applied to all the data symbols and pilot symbols then the OFDM symbol is

$$x(n) = \frac{1}{N}\sum_{n=0}^{N-1} x(k)\, e^{\frac{j2\pi kn}{N}} . \text{for n=1,2,..........N} \qquad (10)$$

The x(k) is either a pilot symbol or a data symbol. At the receiver the received symbol after STBC decoder is applied to FFT block, then the r[th] sub carrier with frequency offset "ε" is given by

$$x_{i,r} = x_{i,r} w_{i,0} + \sum_{r=1, r=r^1}^{N} x_{i,r^1} w_{i,r-r^1} + n_{i,r^1} \tag{11}$$

Where $n_{i,r^1}$ is the noise in the r[th] subcarrier and $w_{i,r-r^1}$ is the weight of ICI coefficient between the rth and r[1]th subcarrier, which is given by[6]

$$w_{i,r-r^1} = sinc(\varepsilon) \cdot \frac{1}{1 + \frac{r - r^1}{\varepsilon}} \tag{12}$$

In eqn (11) the first term $x_{i,r} w_{i,0}$ represents the signal attenuation caused by the frequency offset. When the frequency offset "ε" is null then the ICI coefficient is maximum i.e., $W_{i,(r-0)} = 1$, the second term represents ICI and third is Gaussian noise. The the signal power and ICI power can be estimated from the eqn (11) and it is expressed as

$$P[|C(k)|^2] = E[|x_{i,r} w_{i,0}|^2] \tag{13}$$

$$P[|I(k)|^2] = E[\left| \sum_{r=1, r=r^1}^{N} x_{i,r^1} w_{i,r-r^1} \right|^2] \tag{14}$$

At ICI modulation in transmitter side the data symbols and pilot symbols are mapped on to non neighboring sub carrier's r and (N-r+1) along with their weighting coefficients +1 and -1and these modulated symbols on jth sub carrier can be represented as

$$\begin{aligned} x_{i,j} &= \sum_{r=1}^{N} x_{i,r} w_{i,(r-j)} \\ &= \sum_{r=1}^{\frac{N-N_p}{2}} x_{i,r} \left( w_{i,r-j} - w_{i,N-j-r+1} \right) + \\ &\quad \sum_{l=\frac{(N-N_p+2)}{2}}^{N/2} x_{i,l} \left( w_{i,l-j} - w_{i,N-l-j+1} \right) \end{aligned} \tag{15}$$

In the eqn(15) the second term represents the weighted ICI coefficient at the non neighboring sub carriers and the $N_p$ is the number of pilot symbols added at the receiver side. The sub carriers received with normalized frequency offset "ε" are fed to the ICI demodulation block and these are depicted as

$$Y_{i,j} = X_{i,j^1} - X_{i,N-j+1}$$

(2)

$$- 2X_{i,r}W_{i,0} + \sum_{r=1,r=r^1}^{\frac{N-N_p}{2}} X_{i,r}(W_{i,(r-j)} - W_{i,(N-r-j+1)} - W_{i,(r-N+j-1)}) +$$

$$\sum_{i=\frac{(N-N_p+1)}{2}}^{N/2} X_{i,i}(W_{i,(j-1)} + W_{i,(j-1)} - W_{i,(N-1-j+1)} - W_{i,(1-N+j-1)})$$

(16)

The first term in eqn (16) is the desired signal destroyed by the frequency offset, the second term is the ICI component caused by the data symbols and the last term is another ICI element caused from the pilot symbols. It can be further simplified as

$$Y_{i,j} = 2X_{i,i}W_{i,0} + A + B$$

(17)

The carrier to interference ratio is improved by adding pilot symbols and it can expressed as from[6]

$$CIR = \frac{4|W_{i,0}|^2}{|A+B|^2}$$

(18)

## 3.1 Estimation of CFO

The CFO can be estimated by applying the Maximum Likelihood Estimation (MLE) to two successive OFDM symbols.

1. If the transmitter and receiver is perfectly synchronized the cyclic prefix can be removed perfectly. Then at the receiver after FFT operation the p[th] demodulated pilot symbol of the i[th] OFDM symbol is given by

$$X_{i/p} = \frac{1}{\sqrt{N}} \sum_{n=1}^{N} Y_{i,n} e^{-\frac{j2\pi pn}{N}}$$

(19)

$$X_{i/p} = \frac{1}{N} \sum_{n=1}^{N} \sum_{k=1}^{N} X_{i,k} e^{\frac{j2\pi kn}{N} + \frac{j2\pi \varepsilon n}{N} - \frac{j2\pi pn}{N}}$$

(20)

Where $X_{i,p}$ is a pilot symbol and $Y_{i,n}$ is the received $i^{th}$ OFDM symbol

2. In addition to the $p^{th}$ demodulated pilot symbol $(i+1)^{th}$ OFDM symbol is given by

$$X_{i+1,p} = \frac{1}{\sqrt{N}} \sum_{n=1}^{N} Y_{i+1,n} \, e^{-\frac{j2\pi pn}{N}}$$  **(21)**

It can be elaborated as

$$X_{i+1,p} = \frac{1}{N} \sum_{n=1}^{N} \sum_{k=1}^{N} X_{i,k} \, e^{\frac{j2\pi kn}{N} + \frac{j2\pi\varepsilon(n+N)}{N} - \frac{j2\pi pn}{N}}$$  **(22)**

$$X_{i+1,p} = \frac{1}{N} e^{\frac{j2\pi\varepsilon N}{N}} \sum_{n=1}^{N} \sum_{k=1}^{N} X_{i,k} \, e^{\frac{j2\pi kn}{N} + \frac{j2\pi\varepsilon n}{N} - \frac{j2\pi pn}{N}}$$  **(23)**

$$= X_{i,p} e^{j2\pi\varepsilon}$$  **(24)**

From the eqn(20) and(24) the partial CFO can be estimated as

$$\hat{\varepsilon} = \frac{1}{2\pi} \tan^{-1} \left[ \sum_{p \in \beta} X_{i,p}^{*} \, X_{i+1,p} \right]$$  **(25)**

More accurate partial CFO estimation can be obtained if more pilot symbols are inserted, In addition to that the pilots symbols can be taken on for channel estimation. Besides the efficient performance the pilot symbols decrease the band width utilization. However, the selection of the size and the number of pilot symbols is based on the performance and band width utilization.

## 3.2 Channel models

The channel response depends upon key components like path loss, shadowing, multi path fading, Doppler spread, Co channel and adjacent channel interference. The model parameters are varied according to the atmospheric conditions and these depends upon terrain, tree density, antenna height and beam width and also these parameters are random in nature and only statistical averages are used to characterize them, i.e. in terms of the mean and variance value[8]. Based on all above constraints SUI channels were proposed. There are a set of 6 channels representing 3 terrain types, variety of Doppler spreads, delay spreads and LOS/NLOS conditions.

These channel models can be used for simulations, design, and development and testing of technologies suitable for fixed, broadband wireless applications [7].the following table represents parameters of SUI channel models.

**Table 1.** Classification of SUI channels based on terrain, tree density .

| | | | |
|---|---|---|---|
| Terrain type | C | SUI-1,SUI-2 | Mostly flat terrain with light tree densities. |
| | B | SUI-3,SUI-4 | Hilly terrain with light tree density or flat terrain with moderate to heavy tree density |
| | A | SUI-5,SUI-6 | Hilly terrain with moderate to heavy tree density |

In these channel models the multipath fading is modeled as a tapped delay line with 3taps with non-uniform delays.The gain of the taps and received power are characterized by Rician distribution.

**Table 2.** Various parameters of SUI channels.

| | SUI-1 | SUI-2 | SUI-3 | SUI-4 | SUI-5 | SUI-6 |
|---|---|---|---|---|---|---|
| P(Power in each path in dB) | [0 -15 -20] | [0 -12 -15] | [0 -5 - 10] | [0 -4 -8] | [0 -5 - 10] | [0 -10 -14] |
| K(Ricen Distributi on factor) | [4 0 0] | [2 0 0] | [1 0 0] | [0 0 0] | [0 0 0] | [0 0 0] |
| Tap delay | [0.0 0.4 0.9] | [0.0 0.4 1.1] | [0.0 0.4 0.9] | [0.0 0.5 4.0] | [0 4 10] | [0 14 20] |
| Maximum Doppler frequency (Hz) | [0.4 0.3 0.5] | [0.2 0.15 0.25] | [0.4 0.3 0.5] | [0.2 0.15 0.25] | [2 1.5 2.5] | [0.4 0.3 0.5] |
| Auto_corr (Coefficie nt of antenna Correlatio n) | 0.7 | 0.5 | 0.4 | 0.3 | 0.3 | 0.3 |
| Normaliz ed factor of gain (dB) | -0.1771 | - 0.3930 | - 1.5113 | -1.9218 | -1.5113 | - 0.5683 |

The probability density function of the received power 'r' is given as

$$pdf(r) = \frac{r}{\sigma^2} e^{[-fracr^2+A^2 2\sigma^2]} I_0\left(\frac{rA}{\sigma^2}\right) \quad 0 \leq r \leq \infty \tag{26}$$

$I_0(x)$ represents Bessel function. Here A is representing the LOS/NLOS component and is equal to zero if no LOS component is present. Then the pdf becomes Rayleigh distribution and it is given as

$$pdf(r) = \frac{r}{\sigma^2} e^{[-r^2/2\sigma^2]} \quad 0 \leq r \leq \infty \tag{27}$$

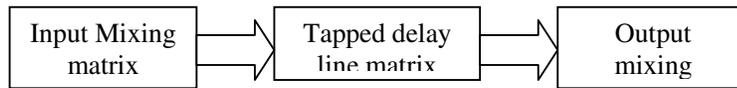

Figure 3. Standard SUI channel model for MIMO-OFDM Systems

## 4. SIMULATIONS

In this section we estimate and compensate the Inter Carrier Interference caused in WiMAX systems. The simulations are carried out on WiMAX system specifications according to IEEE 802.16e [2]. Also considered various channel models of Stanford University Interim (SUI) are considered and these are compared with the ICI cancelling modulation without inserting the pilot symbols. The table specifies the parameters of mobile WiMAX systems.

**Table 3.** Simulation parameters.

| Carrier Frequency | 1.25,5,10,20 GHz |
|---|---|
| Channel Model | LOS/Non-LOS |
| Raw Bit Rate | 1.0-75.0 Mbps |
| Modulation | QPSK, 16QAM, 64QAM |
| OFDM subcarriers | 256 |
| Fading Channel | SUI channel model |
| Guard Interval/Symbol Interval | 1/4, 1/8, 1/16, 1/32 (or 64, 32, 16,8 samples) |
| Frame Duration | 5 ms |
| Number of Frames (per second) | 200 |
| IFFT/FFT | 512 point |
| No.of OFDM symbols in 5ms frame | 48 |
| No. of data sub carriers | 192 |
| No. of pilot sub carriers | 60 |
| Decoder | Viterbi |
| Noise | AWGN |

The Fig .4. shows the estimation of ICI w(r-j) , w^`(r-j) and w^`^`(r-j). It is found that w^`(r-j) < w(r-j) and also the summation in (15) takes only half of the values which leads to the reduction in number of interference signals. Also w^`^`(r-j) will further reduce the cancellation of ICI coefficients and this can be achieved by applying the ICI cancelling modulation.

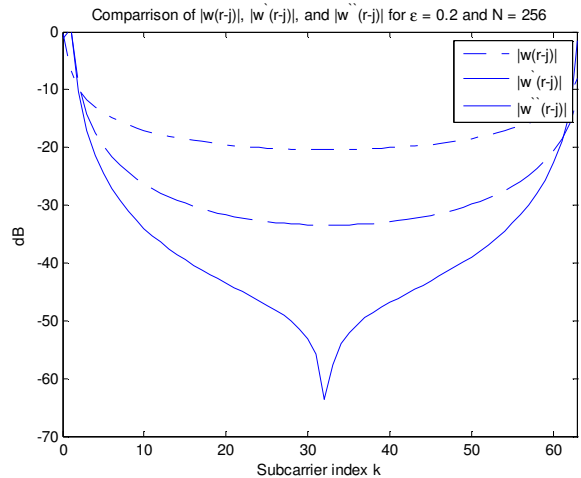

Figure 4.  A comparison between the W(r-j), W^(r-j) and W^^(r-j).

The simulation chain was validated by transmitting data symbols through an ideal channel and by measuring the BER. As expected, the BER in this case is 0, showing that the chain: Randomization, Convolution coding, Interleaver, modulation, Space time encoding, and mapping - OFDM modulation in the presence of Inter carrier interference with a Carrier frequency offset of $\epsilon = 0.2$,under Rayleigh fading channel.

Next, simulations were carried out in all SUI channel models are depicted in Fig .5.

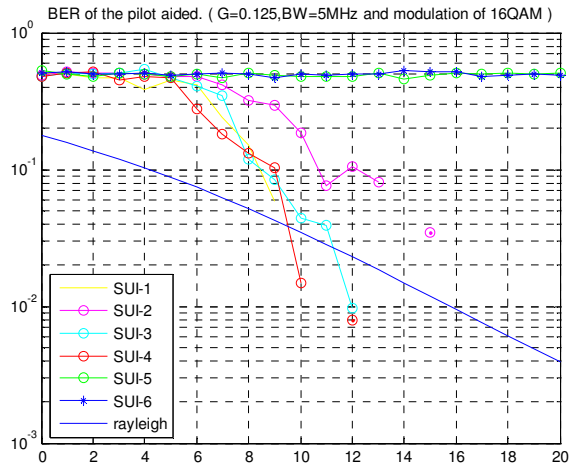

Figure 5.  BER comparison between proposed WiMAX system with SUI channels and Standard WiMAX system with Rayleigh channel.

The simulation results shows the BER for the channel model SUI 1,2,3,4,5 and 6. The BER is high as compared to Rayleigh fading channel because these channel models doesn't have LOS component & due to the impact of ICI the performance is worst. But in case of the channels where the LOS component appears the proposed method performs better and the BER approaches near to a Rayleigh distributed channel. Also as the SNR value increases it may reduce the BER.

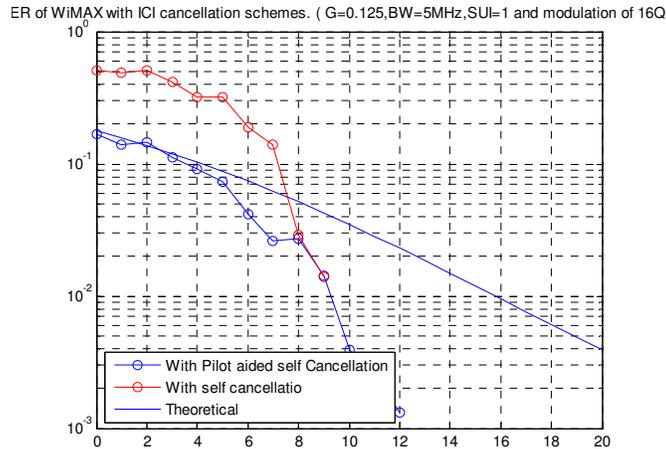

Figure 6.  . BER comparison between WiMAX system with incorporating self cancellation and pilot aided self cancellation withSUI-1 channel

Also the simulations are carried out with another self cancellation scheme[3] and applied to WiMAX and compared with the proposed scheme.  From the Fig.6 we conclude that the proposed method can perform well for the channels where there is an LOS component. The BER reaches to $10^{-3}$ for SNR value at around 12dB as the theoretical Rayleigh channel BER is at almost $10^{-2}$ for the value of SNR at 20dB. In the same way for the same values of SNR the self cancellation method could not obtain less BER as compared to the BER of proposed method with WiMAX.. It is also observed from the simulation carried out on SUI-6 channel that both self cancellation schemes could not attain the BER up to required value. In such situations also the proposed scheme performs well as compared with the self cancellation method.

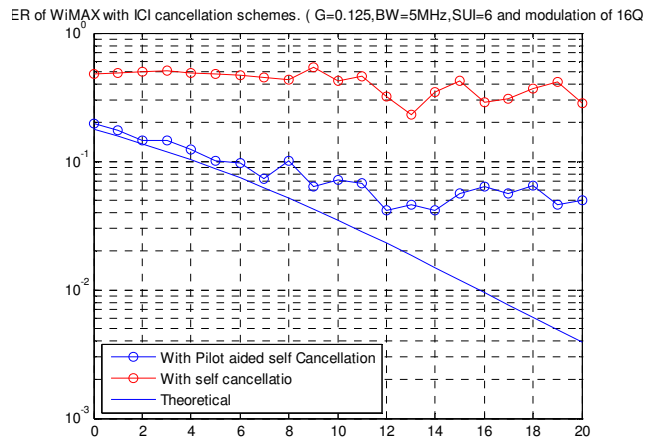

Figure 7.  . BER comparison between WiMAX system with incorporating self cancellation and pilot aided self cancellation.withSUI-6 channel.

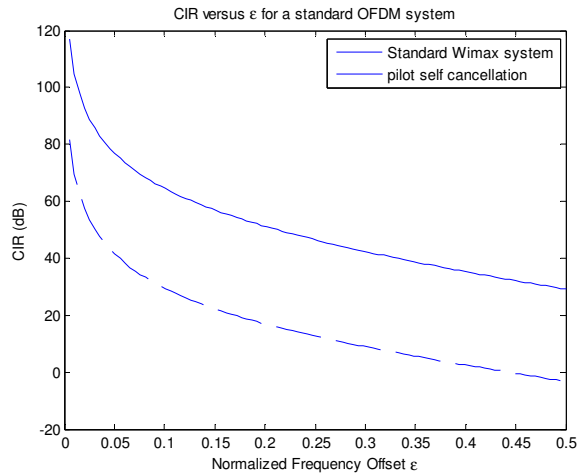

Figure 8.  . CIR comparisons with standard WiMAX system and WiMAX system with Pilot aided self cancellation method.

Finally in WiMAX system theoretically using (18) the proposed method improves the CIR by 4times. By incorporating the ICI PASCS method where as from the fig.8. the CIR increases by 20dB.

## 5. CONCLUSION AND FUTURE SCOPE

In this paper, we propose an Inter Carrier Interference Cancellation method with pilot insertion for WiMAX in various channel environments. The projected scheme provides better performance for Rician distributed channels where LOS component appears in the transmission along with multipath components. The performance of the WiMAX system also improves with the insertion of the proposed method ,and it performs well under different channel conditions with  various characteristics such as Doppler spread, multipath delays  and LOS/NLOS components. But in case of NLOS the performance cannot reach the desired level. Also this scheme may reduce the effective utilization of allotted band. In spite of all these results the Pilot Aided Self Cancellation scheme can be used with IEEE802.16e systems and the performance interms of BER and ICI can be improved further.

**Authors**

**J.TARUN KUMAR** received his B.E degree in Electronics  engineering in 2000 from MIET Gondia, Nagpur University, Maharastra, India. And M.Tech degree in Digital communications from KITS warangal, Kakatiya university, Andhra Pradesh ,India in 2006. He joined in Sumathi reddy Instituteof technology for women in 2011 and currently is Associate professor in the department of electronics and communication engineering. He is also pursuing his Ph.D from JNTUniversity Hyderabad,India .His current research includes 3G enhancements in wireless communications and 4G specifications. He has published 10 papers in National and International Conferences and Jornals.Mr. Tarun kumar is a life member of ISTE, IETE. 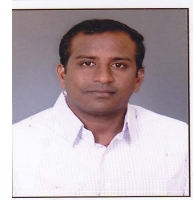

**Dr. K. Anitha Sheela** has done her B. Tech. in Electronics and Communications from REC Warangal during 1989 to 1993 and ME in Systems and Signal Processing from College of Engineering, Osmania University during 1995-1997 and Ph. D from JNTU, Hyderabad during 2003 to 2008. She has worked as Testing Engineer at Onyx industries for 2 years and has been in teaching for more than a decade. She has worked as Coordinator Examination branch, JNTUH and has now taken up the additional responsibility of Additional Controller of Exams apart from her regular teaching profession. She has more than 30 papers published in various National and International Conferences and Journals. Her areas of interest include Speech Processing, Speech Recognition, Speaker Recognition, Pattern Recognition, Image Processing, DSP Processors and Neural Networks. She is also University coordinator for Texas Instruments Embedded Processing Processing Centre established in collaboration with Texas Instruments at ECE Department, JNTUH. She is Fellow of IETE and life member of ISTE. 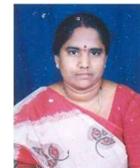